\documentclass[twocolumn,aps,prl]{revtex4}

\usepackage{graphicx}
\usepackage{dcolumn}
\usepackage{bm}
\usepackage{color}
\usepackage{hyperref} 

\begin{document}

\title{Dynamic light diffusion, Anderson localization
 and lasing in disordered inverted opals: \\ 3D ab-initio Maxwell-Bloch computation}

\author{C. Conti$^{1,2}$ and A. Fratalocchi$^{1,2}$}

\affiliation{
$^1$Centro studi e ricerche ``Enrico Fermi,'' Via Panisperna 89/A,
I-00184, Roma, Italy   \\
$^2$Research center Soft INFM-CNR, c/o Universit\`a di
Roma ``La Sapienza,'' I-00185, Roma, Italy }

\date{\today}

\begin{abstract}
We report on 3D time-domain parallel simulations of Anderson localization 
of light in inverted disordered opals displaying a complete
photonic band-gap. We investigate dynamic diffusion processes induced
by femtosecond laser excitations, calculate the diffusion constant and the 
decay-time distribution versus the strength of the disorder. 
We report evidence of the transition from delocalized Bloch oscillations
to strongly localized resonances in self-starting laser processes.
\end{abstract}

\maketitle
The first investigation of strong localization of light 
in periodically structured 3D materials dates back to 
more than twenty years ago \cite{John87}.
Since then the field of {\it photonic crystals} (PhC) has steadily grown, 
and evolved into various directions including all-optical signal processing,
quantum information and bio-sensing \cite{Yablonovitch87,JoannopoulosBook,SakodaBook,SoukoulisBook,Altug06,Lopez06,Noda07,Painter99}.
Notwithstanding these developments, the issue of 3D light localization in the presence
of disorder and backbone periodicity is still largely debated. 
As randomness is unavoidable in any practical realization of PhC, this topic 
has a fundamental importance for the physics and the applications of these devices. 

The starting point is a medium displaying a {\it complete photonic band-gap} (PBG),
i.e. a spectral region where the density of electromagnetic states vanishes.
If disorder is introduced, the PBG is progressively filled by tails of resonances 
at its edges \cite{Li00,Wang03,Yannopapas03};
this circumstance nurtures the observation of 3D Anderson (or strong) localization of light,
which has demonstrated to be elusive in absence of periodicity \cite{Watson87,Storzer06,Wiersma97,Scheffold99,Schuurmans99,Chabanov00}.
Indeed the PBG is filled by states that are expected to be exponentially localized inside the PhC;
following \cite{John87}, these should induce a diffusive
regime of light propagation  \cite{Huang2001,Koenderink05,Sigalas99} 
and a ``critical slowing down of photons'', which results in the development of an exponential 
temporal tail for a pulsed beam propagating in the PhC.
This effect, known as {\it dynamic light diffusion}, is currently largely investigated in disordered systems 
\cite{Storzer06,Chabanov03,Reil05,Skipetrov06,Vellekoop05,Conti07}.

For what concerns 3D PhC, however, there are several open problems that are extremely relevant 
for the final assessment of the observation of Anderson localization of light that is still missing.
Open fundamental questions include the possibility of exciting 
these resonances by short pulses and, correspondingly, 
the expected values of measurable quantities such the dynamical diffusion constant 
and the decay-time distribution of the transmission
(i.e. the way the latter deviates from a simple exponential).

A key issue is the role of disorder-mediated resonances in self-starting lasing (see, e.g, \cite{Cao05}):
if a small percentage of randomness can induce their existence,
any practical light-emitting device based on 3D PhC (as in \cite{Teh07,Scharrer06}) is expected to be strongly affected by
Anderson localizations. Therefore one can argue about their role in the stimulated emission processes if the
 PhC is filled by a light-amplifying material.
{\it Are 3D Anderson localizations able to sustain a laser action? Or delocalized 
``Bloch'' modes always prevail?} 
From a theoretical perspective, the current state of the art is strongly hampered by 
the huge amount of the needed computational resources when dealing with 3D problems.
Indeed simpler low-dimensional models (in 1D and 2D) are quite unsatisfactory for many
important reasons, including the lack of a complete PBG and the critical nature
of Anderson localizations in 3D \cite{ShengBook}.

In this manuscript, we apply a massively parallel approach to give answers to the mentioned open
problems. For a realistic finite structure, we show the way the PBG is closed when increasing the strength of the disorder, 
we provide evidence of the excitation of localized modes by using short pulses, and we determine the relation between
the diffusion constant $D$ and the strength of the disorder. We find that, 
in the strongly localized regimes (when $D$ vanishes), the decay-time distribution
of the transmission splits into a fast and a slow component,
showing a clear dynamical signature of the 
existence of a second regime of light localization in disordered PhC.
Finally, in order to ascertain if Anderson resonances are actually involved in lasing processes,
we solve from first principles the most general semi-classical model of light-matter interaction to simulate
a self-starting laser oscillation. Specifically, we show the transition from emission mediated by delocalized
periodical waves (Bloch modes) to that driven by Anderson localization.

\begin{figure}
\includegraphics[width=8.3cm]{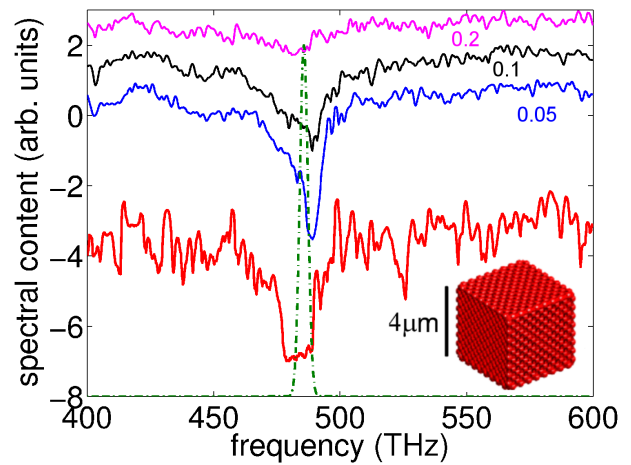}
\caption{ (Color online)
Spectral content (log. scale, each line is vertically shifted
by an arbitrary amount) 
of the electromagnetic field inside the
3D inverted opal for increasing degree of disorder;
the thick line is the ordered case ($\gamma=0$).
The dot-dashed line is the spectrum (arbitrary linear scale)
of the pulsed $100\text{fs}$ excitation for the dynamic diffusion analysis (see text).
Inset: sketch of the 3D PhC index profile.
\label{figlinear}}
\end{figure}

{\it Sample and numerical approach ---}
Our approach is based on a parallel Finite-Difference Time-Domain (FDTD) code \cite{TafloveBook}
with typical runs employing from tens to hundreds of processors on Linux clusters
and IBM-SP5 systems. 
With reference to lasing processes, we solve the Maxwell-Bloch equations for a four-level system
within a first-principle formulation \cite{Conti07Mie}. 
At variance with previously reported 1D and 2D models  \cite{Jiang00,Vanneste05,tl_fd_slavca},
three levels are taken as degenerated and coupled 
to the three components of the electric field, thus accounting
for a fully vectorial representation of the light interaction
with a resonant system.
The density matrix equations for a four level systems are re-written in terms of real-valued fields
by applying an homomorphism with the su(4) algebra \cite{Conti07Mie} and solved in the
middle of the Yee grid. 
The parameters are chosen in order to mimic a typical dye, i.e. Rhodamine B,
with resonance wavelength  $484\text{THz}$ ($\lambda_0=620\text{nm}$) and refractive index $n=1.3$. 
The characteristic time-scales for the density matrix equations are set to $T_1=1$~ns and $T_2=10$~fs, 
providing a small signal gain bandwidth of the order of $1/T_2=100T$Hz.
The noise due to spontaneous emission is included as a stochastic source term in the electric field evolution \cite{tl_fd_slavca}.
The integration domain is a box with size $L_{box}=6\mu\text{m}$ and
``UPML'' boundary conditions \cite{TafloveBook}.
Most of the reported simulations adopt grid sizes of $200^3$, with $400^3$ high-resolution runs for checking. 
The considered finite-size PhC (edge $L=4\mu\text{m}$) is an inverted opal (Fig.\ref{figlinear} inset)
with dielectric constant is $\epsilon_r=12$,  displaying a complete PBG around $a/\lambda=0.8$
where $a$ is the cubic lattice constant.
In absence of disorder, the radius of each sphere in the PhC is $r_0=a/2\sqrt{2}$  \cite{Ho90,Busch98, Blanco00,Sozuer92,Lodahl2004,Andreani07}.
Results are reported with reference to the case $a=480\text{nm}$,
providing a band gap around $500\text{THz}$ ($\lambda=600nm$). 
Disorder is finally introduced by a random-number generator providing
uniform deviates $\xi$ in $[-1/2,1/2]$, and such that the radius
of each air sphere in the inverted opal is given by $r=r_0(1+\gamma \xi)$,
where $\gamma$ measures the strength of the disorder
($\gamma=0$ is the ordered case).

{\it Wide-band excitation and gap filling.}
In the first set of runs we consider the passive air-sphere PhC
and investigate the way the forbidden band 
is closed when increasing the strength of the disorder.
We launch a single cycle (linearly polarized $TEM_{00}$ with waist $2\mu$m) pulse 
with frequency  $484\text{THz}$ ($\lambda=617.25nm$) in the [100] crystallographic direction
(denoted as $z$ hereafter), and we retrieve the field in a low symmetry point inside
the sample. After a total time interval of $2.9ps$ ($10^5$ time-steps with
$200^3$ grid and $dt\cong2.9\times10^{-17}$s), we calculate the electromagnetic oscillation spectrum 
with resolution $\sim0.4$~THz.
Figure \ref{figlinear} shows the field spectrum for various $\gamma$, 
with results averaged over $10$ realizations of the disorder
(the total processor time for figure \ref{figlinear} is of the order of $480$ days).
For large $\gamma$ the PBG
is closed by tails of states
at the band-edges. This demonstrates that of a few percent of randomness is sufficient to close the 
forbidden band.

\begin{figure}
\includegraphics[width=8.5cm]{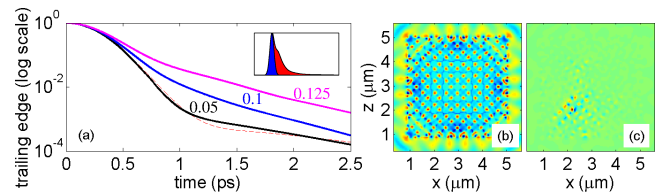}
\caption{(Color online) 
(a) 
Tail of the transmitted pulse for various $\gamma$ (semi-log. scale, dashed line for $\gamma=0$), the
time axis has been shifted such that the peak of $T(t)$ corresponds to $t=0$;
the inset shows the transmitted pulse for $\gamma=0.1$ (light region, total time window $2ps$) with
the input pulse superimposed (dark region);
(b) Snapshot of $E_y$ component in the PhC middle $(y,z)$ section for $\gamma=0$ and $t\cong 6 ps$;
(c) as in (b) for $\gamma=0.1$.
\label{figpoynting}}
\end{figure}

{\it Dynamical diffusion and localization.}
In a second set of runs we consider the dynamic light diffusion 
process of a $100fs$ Gaussian pulse with spectrum located within the PBG
in proximity of the the high-frequency edge (Fig.\ref{figlinear},
other beam parameters are the same as the single-cycle case considered above).
We record the transmitted power pulse $T(t)$ (transversely integrated Poynting vector at the output face of the PhC as in \cite{Conti07}). 
Figure \ref{figpoynting}a shows $T(t)$ for increasing $\gamma$:
we observe the development of a pronounced temporal tail, which   
unveils the ``critical slowing down of photons'' \cite{John87};
this is also evident in the inset \ref{figpoynting}a by comparing the input pulse
with the transmitted one in the case $\gamma=0.1$.
Figure \ref{figpoynting}b,c show
the snapshot of the $E_y$ component of the electric field
in the middle $x$-section plane $(y,z)$ taken at very large times from the peak value of the input pulse.
(in this way we select the long living modes in the 
structure).
Note that in the absence of disorder, (Fig.\ref{figpoynting}b)
a delocalized periodical Bloch mode lying in at the band-edge survives;
it corresponds to the high-frequency spectral tails 
of the input pulse. Time-domain spectral analysis (not reported) furnishes
 its frequency $f\cong490$THz ($\lambda\cong614nm$).
In the presence of disorder $\gamma=0.1$ (Fig.\ref{figpoynting}c),
the field snapshot at long times reveals 
the onset of a localized states \cite{John87} that turns out to correspond to $f\cong484$THz ($\lambda\cong620$nm).

\begin{figure}
\includegraphics[width=8.3cm]{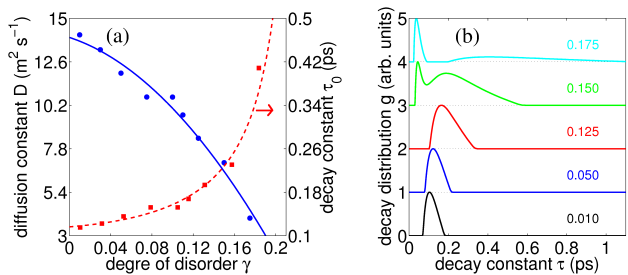}
\caption{(Color online) 
(a) 
Diffusion constant $D$ (dots, left-scale), quadratic polynomial best-fit (continuous line), 
and time delay $\tau_0$ (squares and best-fit dashed line, right scale).
(b) Decay-time distribution $g(\tau)$ for various $\gamma$,
all the lines have been vertically scaled to the peak value.
\label{figdiffusion}}
\end{figure}

{\it Diffusion constant and decay times.}
Figure \ref{figdiffusion}a shows the dynamic light diffusion constant $D$ versus the degree 
of disorder. We calculate $D$ by fitting the trailing edge of $T(t)$ by an exponential tail $\exp(-t/\tau_0)$,
where $\tau_0$ is the leading time constant; $D$ is then calculated by $D=L^2/\pi^2 \tau_0$. A quadratic fit for $D$ gives 
\begin{equation}
D(m^2 s^{-1})\cong 14-19\gamma-204\gamma^2\text{.}
\label{diffusioneq}
\end{equation}
Equation (\ref{diffusioneq}) implies that $D=0$ when $\gamma\cong0.2$.
At that value for $\gamma$ the PBG is closed (Fig.\ref{figlinear})
and a simple exponential fit for the tail of $T(t)$ does not work.
We study this regime by the distribution $g(\tau)$ of the decay-times (Fig.\ref{figdiffusion}b).
We find $g(\tau)$ by minimizing the difference between the trailing edge of the 
transmission $T(t)$ and a signal given by the sum of 
exponential functions with decay-constant $\tau$ and weight $g(\tau)$.
While increasing $\gamma$, $g(\tau)$ spreads
around the leading diffusion time $\tau_0$.
Therefore, $T(t)$ progressively deviates from a simple exponential 
unveiling the excitation of localized states.
Notably enough, as $\gamma\gtrsim 0.15$ the time constant distribution is splits into a fast and a slow component (Fig.\ref{figdiffusion}b).
This behavior is due to the fact that when the PBG is closed, some of the input energy is 
trapped in the localized states [thus providing the long time tails in $T(t)$], while other
travels by tunneling through low-Q factor modes, which do not exist at small $\gamma$.
Previous investigations on low-index contrast PhC \cite{Vlasov99} envisaged the existence of a
second localization regime in a complete PBG in the presence of large disorder.
Our analysis shows that there is a transition in the decay-time distribution in correspondence
of the zero of the diffusion $D$, when the PBG is practically filled 
by tails of localized states.
This is an experimentally testable signature of a specific regime of light localization,
which appears as a characteristic feature of PhC.

{\it 3+1D simulation of lasing in disordered and ordered PhC.}
\begin{figure}
\includegraphics[width=8.3cm]{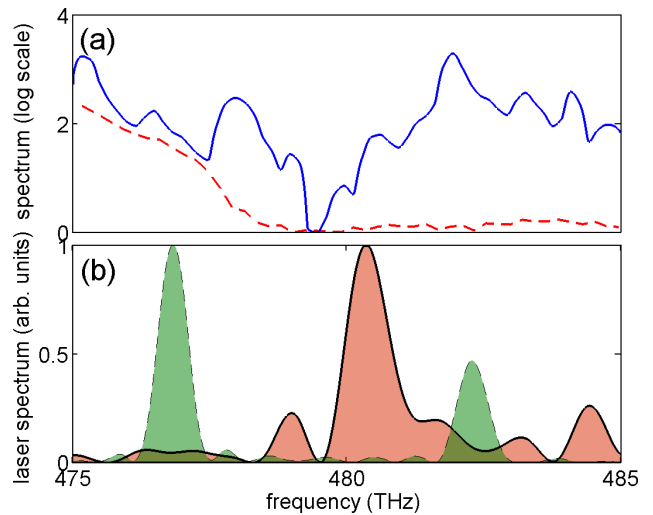}
\caption{(Color online) 
(a) Wide-band single-cycle excitation spectrum (log. scale) for $\gamma=0$:
filled (continuous line) and the empty opal (dashed line).
The spectra have been vertically translated of an arbitrary amount for
the sake of comparison.
(b) Self-starting laser spectrum in the ordered (dashed thin line)
 and in the disordered case (thick line) ($\gamma=0.1$).
\label{figspectrumlaser}}
\end{figure}
To model laser emission in the PhC, we take the air-spheres as filled by an active resonant medium (modeled
by the Maxwell-Bloch equations detailed above).
We first analyze the way the 
presence of the infiltrated medium affects the width of the forbidden-gap.
Figure \ref{figspectrumlaser} compares the responses of an air-filled (dashed-line) and dye-filled (thick-line) opal
to a single-cycle pulse. The forbidden-band is progressively closed as the refractive index of the infiltrated
medium increases; indeed we found that a ``small'' gap survives around $f=480\text{THz}$ ($\lambda=625\text{nm}$).
We then consider the case the active medium is externally pumped.
At low inversion $N_a$, 
the laser is below threshold and only fluorescence is obtained (not reported).
At high pumping ($N_a=5\times 10^{25}$~m$^{-3}$), in the ordered case ($\gamma=0$), we find that the PhC cavity starts oscillating on
a spectrum mainly concentrated at the band-edges of the linear PBG of the infiltrated opal
(see Fig. \ref{figspectrumlaser}b). This spectrum corresponds to the excitation of Bloch-modes,
 whose spatio-temporal profile is delocalized. This is confirmed by Fig.\ref{fig3d}a, showing a 3D iso-surface plot of the
electromagnetic energy density stored inside the structure at long times ($t\cong 3$~ps) since the turning on of the pump.
Conversely, in the disordered case ($\gamma=0.1$), we find that the emission spectra is mostly concentrated
within the linear band gap  of the filled case (Fig. \ref{figspectrumlaser}).
The corresponding mode profile is clearly localized, as shown in the panel (b) in Fig.\ref{fig3d}.
\begin{figure}
\includegraphics[width=8.3cm]{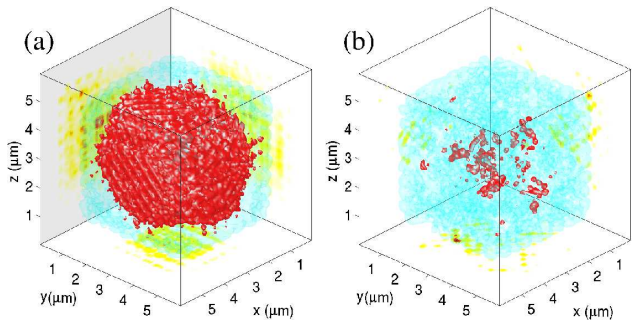}
\caption{(Color online) 
3D iso-surface for the electromagnetic energy density at high pumping rate: (a) ordered $\gamma=0$ and (b) disordered case $\gamma=0.1$.
Opaque surface corresponds to the energy density level of $3\,\text{fJ}/\mu\text{m}^3$; 
transparent surface corresponds to $0.2\,\text{fJ}/\mu\text{m}^{3}$;
the lateral plots are projection of the middle sections of the energy density.
\label{fig3d}}
\end{figure}

{\it Conclusions ---} We reported on an extensive investigation
of the interplay between localization and disorder 
in an inverted opal exhibiting a complete photonic band gap.
The dynamic diffusion regime in response to an in-gap $100$fs
light pulse excitation shows that the diffusion constant decreases when increasing
the strength of the disordered and localized modes are put into oscillation.
The time-dependent transmission develops a non-exponential
tail, evidencing of a second regime characterized by the splitting of the decay-time distribution at large disorder.
The fully vectorial 3D Maxwell-Bloch simulation of self-starting lasing processes
finally demonstrates that, in the presence of disorder,  delocalized Bloch-modes are replaced by the Anderson localizations in sustaining stimulated emission processes.

{\it Acknowledgments ---}
We acknowledge discussions with G. Ruocco, C. Toninelli and D. Wiersma, and
 support from the INFM-CINECA initiative for parallel computing.


\end{document}